# Pancakification and negative Hawking temperatures

Tyler McMaken

*JILA and Department of Physics, University of Colorado,
Boulder, Colorado 80309, USA*

tyler.mcmaken@colorado.edu

Dated 29 March 2023

**Abstract:**

Vacuum models of charged or spinning black holes possess two horizons, the inner of which has the oft-overlooked property that gravitational tidal forces initially spaghettifying a freely falling observer will eventually change signs and flatten the observer like a pancake. Inner horizons also induce a classical blueshift instability known as mass inflation, and a number of recent studies have found that inner horizons exhibit even stronger quantum singular behavior. In this essay we explore the quantum effect of Hawking radiation, which in the presence of compressive tidal forces seems to predict negative temperatures. By analyzing the interaction of quantum fields with black hole geometries, we can come to a closer semiclassical understanding of what really happens near a black hole's inner horizon.

What happens when an observer falls into an astrophysical black hole? A typical answer is the following: as they pass through the (locally undetectable) event horizon, gravitational tidal forces begin to stretch them out radially like a noodle and squash them in the transverse directions in a process known as spaghettification. Eventually, these forces will exceed their body's natural tensile strength, they will be ripped apart, and what remains will continue to fall inwards until unavoidably hitting a spacelike singularity, where all known laws of physics break down.

Such a story is technically only true for the simplest possible black hole model, the Schwarzschild metric. For more astrophysically relevant black hole models that possess non-zero angular momentum, outward centrifugal forces will begin to slow the inflow of space [1] close to the singularity, forming a second horizon deeper inside the black hole known as the inner (Cauchy) horizon. As an observer approaches this inner horizon feet-first, the pull of gravity at their feet will become *weaker* than the pull of gravity at their head, and instead of being spaghettified, they will be flattened out like a pancake.

Tidal forces in general relativity can be described using the Jacobi vector field $\eta(\tau) \equiv \partial \gamma_s(\tau)/\partial s$ for a smooth, one-parameter family of closely spaced geodesics $\gamma_s$ with affine time $\tau$. In terms of observer coordinates $x^\mu$, the equations for geodesic deviation are [2]

$$\frac{D^2 \eta^\mu}{D\tau^2} = R^\mu{}_{\nu\rho\sigma} \dot{x}^\nu \dot{x}^\rho \eta^\sigma, \qquad (1)$$

where $R^\mu{}_{\nu\rho\sigma}$ is the Riemann curvature tensor, $D$ denotes directional covariant differentiation, and an overdot denotes total differentiation with respect to $\tau$.

Consider the Reissner-Nordström (RN) metric for an electrically charged black hole:

$$ds^2 = -\Delta(r)\, dt^2 + \frac{dr^2}{\Delta(r)} + r^2(d\theta^2 + \sin^2\theta\, d\phi^2), \qquad (2)$$

where the radial function

$$\Delta(r) \equiv \left(1 - \frac{r_+}{r}\right)\left(1 - \frac{r_-}{r}\right), \qquad r_\pm \equiv M \pm \sqrt{M^2 - Q^2}, \tag{3}$$

which is often called the horizon function, contains zeros at the positions of the outer ($r_+$) and inner ($r_-$) horizons. While astrophysical black holes are not expected to possess substantial electric charge, the RN metric has the same causal structure as spinning black holes but with more symmetries—in the words of John Wheeler, "charge is a poor man's angular momentum."

For a RN observer in radial free fall, the geodesic deviation equations simplify to [3]

$$\frac{\ddot\eta^r}{\eta^r} = -\frac{\Delta''}{2}, \qquad \frac{\ddot\eta^\theta}{\eta^\theta} = \frac{\ddot\eta^\phi}{\eta^\phi} = -\frac{\Delta'}{2r}, \tag{4}$$

where a prime denotes differentiation with respect to $r$. In the Schwarzschild case (with charge-to-mass ratio $Q/M = 0$), the radial geodesic deviation $-\Delta''/2$ is always positive, while the angular deviations $-\Delta'/(2r)$ are always negative, as expected for radial tidal stretching and transversal tidal compression. But when the black hole's charge is non-zero, both the radial and transversal tidal accelerations reverse as $r$ is decreased to $r_-$, so that a free-falling observer will first be spaghettified, then compressed in all directions as the radial tidal force changes signs, and finally "pancakified" as the angular components switch signs. These angular tidal forces always switch signs beneath the event horizon, but the radial tidal acceleration can become negative above the event horizon for a black hole near enough to extremality; in particular, when $(Q/M)^2 > 8/9$.

Why care about tidal stretching and compression in black hole spacetimes? Aside from potential astrophysical implications for near-horizon accretion physics, tidal forces are often used as a heuristic picture for the production of Hawking radiation, in which virtual pairs of particles emerging from the vacuum are ripped apart near the event horizon so that one escapes to infinity and the other is captured by the black hole (such a picture was first proposed by Hawking himself in his 1975 article [4] and formalized decades later by Parikh & Wilczek [5]). One can imagine a gravitational analog to the Schwinger effect [6] (the creation of electron-positron pairs from the QED vacuum), with one complication: whereas electric fields act in opposite ways for charges of different signs, gravity always acts attractively on positive rest masses. Energy conservation can thus only be satisfied in the presence of a horizon (or more accurately, an ergoregion), where particles can exist with negative energy as measured by an observer at rest at infinity.

While Hawking radiation is typically pictured as originating at or just above the event horizon, virtual pairs are being ripped apart everywhere within the ergoregion and can be detected whenever the appropriate observer (e.g., a two-level atom) presents themself. In fact, one of the key insights we wish to convey here is that the miniscule levels of Hawking radiation that manage to leak out to infinity are nothing compared to the roiling atmosphere of quantum radiation near the inner horizon. However, one encounters quite a few problems when taking the local, heuristic picture of Hawking radiation literally—for example, for black holes near enough to extremality, virtual pairs near the event horizon will not be ripped apart by tidal forces; instead, they will be compressed together in all directions. What, then, if anything, would an infaller locally observe?

Thus, let us turn to a more rigorous formalism for the local perception of Hawking radiation from an infaller. Instead of particles, Hawking imagined waves within a quantum field, originating from a Minkowski vacuum in the asymptotic past, scattering through the dynamical formation of a black hole, and emerging at future null infinity in a different vacuum state [4]. The expectation value of the number of excited modes in the *out* region from the vacuum state of the *in* region is governed by the Klein-Gordon inner product of these wave modes $\Phi_{\omega\ell m}$ from the two regions, via the Bogoliubov coefficient $\beta_{\omega\ell m}^{\bar\omega\bar\ell\bar m} = \langle\Phi_{\bar\omega\bar\ell\bar m}^{in}|(\Phi_{\omega\ell m}^{out})^*\rangle$:

$$\langle 0_{in}|\hat{N}_{out}|0_{in}\rangle = \int_0^\infty d\bar\omega \sum_{\bar\ell=0}^\infty \sum_{\bar m=-\bar\ell}^{\bar\ell} |\beta_{\omega\ell m}^{\bar\omega\bar\ell\bar m}|^2. \tag{5}$$

If these wave modes $\Phi^{out}_{\omega \ell m}$ and $\Phi^{in}_{\bar\omega \bar\ell \bar m}$ asymptotically behave sinusoidally in null coordinates $u$ and $U$ (respectively) that satisfy the exponential peeling relation

$$U = U_H - A\, e^{-\kappa u} \tag{6}$$

for some constants $\kappa$, $U_H$, and $A$, then the inner product of these modes will yield a thermal spectrum with temperature $\kappa/(2\pi)$ [4]:

$$|\beta^{\bar\omega \bar\ell \bar m}_{\omega \ell m}|^2 \propto \frac{1}{e^{2\pi\omega/\kappa} - 1}. \tag{7}$$

While Hawking's original formalism applies only to modes reaching future null infinity, one can easily generalize so that the *out* state is replaced by the vacuum state in the locally inertial tetrad frame of any infalling observer at any spacetime position. A few mild technical assumptions are warranted [7], but the result is that if $\kappa$ is replaced by a function

$$\kappa_{\text{eff}}(u) \equiv -\frac{d\ln(dU/du)}{du} \tag{8}$$

that remains sufficiently adiabatic around the local neighborhood of an observer at $u$ (via the condition $|d\kappa_{\text{eff}}/du| \ll \kappa^2_{\text{eff}}$), then a thermal Hawking flux will be observed with a temperature $\kappa_{\text{eff}}(u)/2\pi$. This effective temperature function $\kappa_{\text{eff}}$ then reduces to the exact surface gravity $\kappa$ when the observer is taken asymptotically far from the black hole.

The calculation of $\kappa_{\text{eff}}(u)$ for various families of observers in the Schwarzschild spacetime can be done by explicitly finding the relation between $u$ and $U$ [8]; however, for more complicated spacetimes, it will be useful to approach the problem from a different angle. Since $dU/du$ represents variations in the proper times of two observers (one to define the past vacuum state and the other to detect the radiation) connected by a null geodesic, Eq. (8) can be recast into the form

$$\kappa_{\text{eff}}(\tau_{\text{ob}}) = -\frac{d\ln(\omega_{\text{ob}}/\omega_{\text{em}})}{d\tau_{\text{ob}}}, \tag{9}$$

where the frequency $\omega \equiv -k^\mu \dot{x}_\mu$ is the timelike component of a null particle's 4-velocity $k^\mu$ measured in the frame of an *in* state emitter (subscript "em") or *out* state observer (subscript "ob") with 4-velocity $\dot{x}^\mu$. This quantity can be straightforwardly calculated for any observer in any spacetime, and further, it can be generalized from the radial case to any direction in the observer's field of view (with, as one might expect, some mild technical assumptions [9]).

How does $\kappa_{\text{eff}}(\tau_{\text{ob}})$ behave for a freely falling observer in the RN spacetime? (The same arguments as those given below also hold true for rotating black holes but require additional non-important complications [10].) When the observer looks in the radial directions (let us call the effective temperature $\kappa_{\text{eff}}^+$ when looking downwards at outgoing modes and $\kappa_{\text{eff}}^-$ when looking upwards at ingoing modes), the effective temperature can be written simply as [11]

$$\kappa_{\text{eff}}^\pm(r_{\text{ob}}) = \mp \frac{\omega(r_{\text{ob}})}{\omega(\infty)} \left( \frac{\Delta'(r_{\text{ob}})}{2} - \frac{\Delta'(r_{\text{em}})}{2} \right), \tag{10}$$

where the emitter defining the *in* state is located at $r_{\text{em}} \to r_+$ for outgoing modes and $r_{\text{em}} \to \infty$ for ingoing modes; this is the so-called Unruh vacuum state. The effective temperature seen by an inertial observer is thus given by a Doppler frequency factor multiplied by two terms, an observer-dependent surface gravity $\Delta'(r_{\text{ob}})/2$ and a state-dependent surface gravity $\Delta'(r_{\text{em}})/2$. When the observer's radius $r_{\text{ob}}$ is taken to infinity, the observer's surface gravity vanishes, and all that remains when looking downward is the surface gravity $\kappa = \Delta'(r_+)/2$ predicted by Hawking.

One might assume from inspecting Eq. (10) that an observer passing through the event horizon and looking downward would see no radiation, since the two terms in parentheses appear identical. However, at the horizons ($r_{\text{ob}} \to r_\pm$), the Doppler term $\omega(r_{\text{ob}})/\omega(\infty)$ diverges as $\Delta^{-1}$, and thus a more careful calculation is needed. Focusing on a positive-energy ($E_{\text{ob}}$), free-falling observer looking downward ($r_{\text{em}} \to r_+$), the effective temperature seen at the event horizon is [10]

$$\kappa_{\text{eff}}^+(r_+) = -E_{\text{ob}} \frac{\Delta''(r_+)}{\Delta'(r_+)}. \tag{11}$$

Coincidentally enough, this effective temperature has the same qualitative behavior as the tidal forces of Eq. (4) at the event horizon: $\kappa_{\text{eff}}^+(r_+)$ is usually small and positive (just as it is at infinity), but for a black hole near enough to extremality, and in fact precisely when $(Q/M)^2 > 8/9$, the effective temperature becomes negative.

One can easily verify that the situation worsens when an observer approaches the inner horizon. There the surface gravity terms of Eq. (10) no longer cancel, and as a result, the blueshift-diverging Doppler term $\omega(r_{\text{ob}})/\omega(\infty)$ will cause the now-negative $\kappa_{\text{eff}}^+$ to diverge as $\Delta^{-1}$ at the inner horizon (while $\kappa_{\text{eff}}^-$ remains finite and negative there). As it turns out, this semiclassical divergence occurs for *any* black hole with separated inner and outer horizons, even if the interior is regularized to remove singularities or mass inflation instabilities [10]. An observer falling into a black hole will be met with an increasingly dense atmosphere of Hawking radiation as they plunge through the interior until they reach a wall of infinite energy at the inner horizon.

Can these results be trusted? What does it even mean to have a negative temperature, and what about dynamical back-reaction? One may suspect that the adiabaticity condition is not satisfied at the inner horizon and therefore that the effective temperature defined there is meaningless. As it turns out, the quantity $|\dot{\kappa}_{\text{eff}}^+|/(\kappa_{\text{eff}}^+)^2$ controlling the degree of adiabaticity is always less than unity at the inner horizon, so the Hawking temperature should be approximately thermal there (at least in the geometric optics limit) [11]. But to confirm that the semiclassical inner horizon divergence can be trusted, one may compute the full spectrum numerically from Eq. (5) using scattering theory. Such a calculation was carried out for a large range of black hole parameters in [11], all yielding the same qualitative results, as shown in Figure 1.

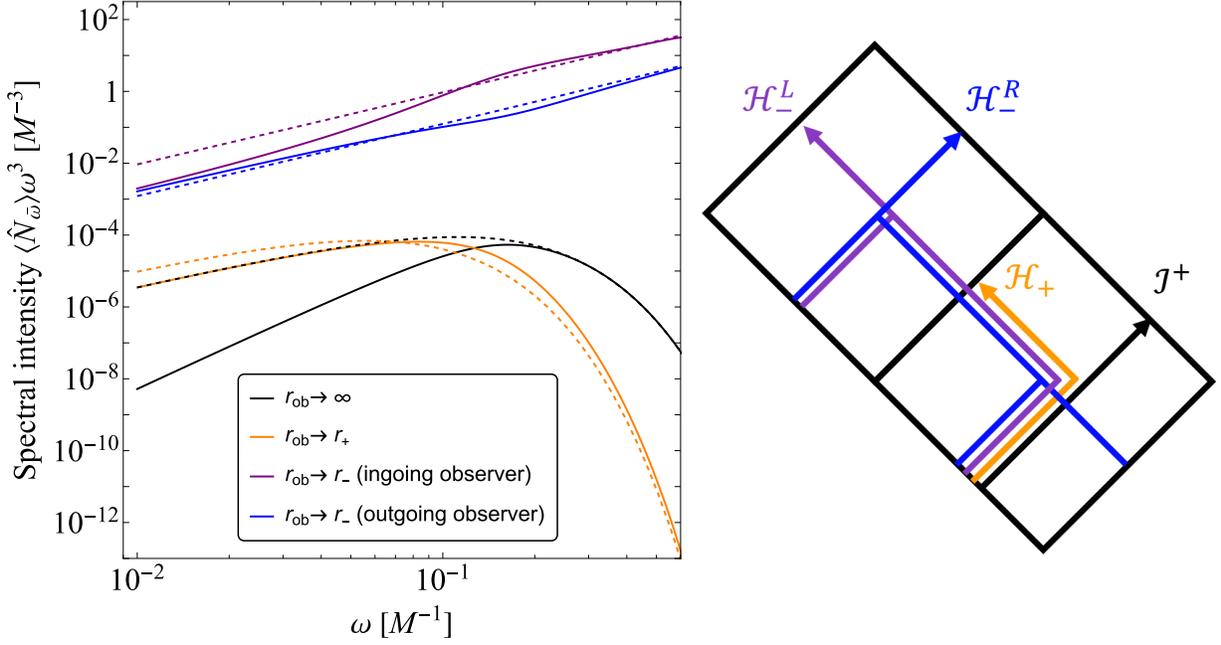

*Figure 1.* (Left panel) Spectral intensity of Hawking radiation seen by an observer at various positions within a Reissner-Nordström spacetime with $Q/M = 0.5$. Solid curves show numerically evaluated spectra from Eq. (5), and dashed curves show the corresponding thermal spectra at temperatures $\kappa_{\text{eff}}(r_{\text{ob}})/(2\pi)$ from Eq. (10). While the spectra seen at infinity and the event horizon yield standard graybody deviations from a Planckian distribution, the spectra seen at the inner horizon are much greater in intensity and diverge in the ultraviolet limit. (Right panel) Penrose diagram showing the null paths contributing to the scattered modes used to compute the spectra.

For an observer at infinity (black curves in Figure 1), the spectrum of Hawking radiation looks approximately like a blackbody, with a Rayleigh-Jeans power law at lower frequencies and an exponential drop-off at higher frequencies. The exact blackbody predicted by the effective temperature $\kappa_{\text{eff}}^+$ is shown by the dashed line, which coincides with the numerically evaluated Bogoliubov coefficient in the geometric optics (high frequency) limit. But at low frequencies, the graybody spectrum differs by a power law of 2, as first predicted by Starobinsky & Churilov [12].

When the observer approaches the event horizon, the Hawking spectrum still appears approximately as a blackbody, with a lower temperature than that at infinity (by a factor of 2, in this case). But once the observer approaches the inner horizon, the spectrum no longer appears Planckian. Instead, the exponential drop-off at high frequencies inverts, leading to an ultraviolet divergence. As shown by the dashed curves, these spectra approximately match the magnitudes one would obtain for a Planckian distribution with *negative* temperatures $\kappa_{\text{eff}}^-/(2\pi)$ and $\kappa_{\text{eff}}^+/(2\pi)$ at the left and right inner horizons, respectively (note that while $\kappa_{\text{eff}}^+$ diverges at the left inner horizon, $\kappa_{\text{eff}}^-$ remains finite and negative there, and similarly for the right inner horizon). The inner horizon spectra shown in Figure 1 thus are the tamer cases compared to the much more violent divergences seen when the observer looks in the opposite direction.

By now it should be clear that stationary vacuum black hole metrics with inner horizons must be modified drastically when semiclassical effects are taken into account. They will become semiclassically singular at the inner horizon. Such a singularity is not only seen in the perception of Hawking radiation presented here, but it is also seen with a quantized scalar field's renormalized stress-energy tensor, as has been found in a monumental series of recent works at the beginning of the global pandemic by several different groups [13–16]. Even if a black hole is perfectly isolated and classically stable to perturbations, the mere existence of a quantum vacuum will act to close off the wormhole hiding behind the Cauchy horizon.

The entire discussion above makes the assumption that the background metric remains fixed in the presence of a quantum field. Will the inner horizon singularity persist if dynamical evolution is permitted? Classically, back-reactions from perturbations tend to push the inner horizon inward until it is forced to stop at $r = 0$ and form a spacelike singularity [17]. But semiclassically, in the *s*-wave approximation, the negative temperature will tend to push the inner horizon outward at an extremely rapid rate until it reaches the event horizon and forms an extremal or compact horizonless object [16]. Clearly both classical and semiclassical effects must be taken into account in a self-consistent manner to understand how astrophysical black holes truly evolve.

One potential path forward involves a classical perturbation model of mass inflation from Hamilton and collaborators [18–20]. Instead of considering perturbations from discrete collapsing null shells, one can ask what the most general solution to Einstein's equations is for a black hole that, like the Kerr metric, is axisymmetric and (conformally) Hamilton-Jacobi separable, but, unlike the vacuum case, contains a steady conformal accretion of matter. Such a metric may not be valid on cosmological time scales into the future (just as the asymptotically flat Kerr metric is not), but in the present cosmological era, black holes are continually accreting light and matter (even during their quiescent phases) that are orders of magnitude more intense than the Price trail of gravitational radiation typically used to assess perturbation dynamics. The resulting metric predicts that the black hole will appear practically identical to the Kerr metric everywhere in the exterior, but close to the inner horizon, the geometry will collapse to form a spacelike singularity. An infalling observer there will undergo a series of BKL-like bounces, first being pancakified in a manner akin to the tidal process described at the beginning of this essay, but then once again being spaghettified as hyper-relativistic streams of matter build up radially in both directions.

The solution for Hamilton's conformally accreting, rotating black hole is far too complex to make an analytic calculation of the semiclassical back-reaction feasible, but close to the inner horizon, the equations simplify considerably [21]. In this so-called inflationary Kasner regime, the same divergence of Hawking radiation as described above appears at the classical spacelike singularity, and though the full renormalized stress-energy tensor governing the quantum back-reaction remains to be computed, calculations of the building blocks of this quantity indicate that the quantum stress-energy contribution should take the same form as the classical contribution [22]. What this implies is that the metric is semiclassically self-consistent—both classical and quantum perturbations would feed back into the same inflationary mechanism causing the geometry to collapse into a spacelike singularity.

When a quantum field interacts with an astrophysical black hole geometry to first order, one should thus never expect an inner horizon to survive. While an observer should semiclassically be able to pass through the event horizon with no worries except a bit of spaghettification, their continued journey will be met with mayhem as they dive within the quantum atmosphere, become pancakified, and potentially become spaghettified again as the spacetime back-reacts to form a stable, spacelike singularity. Once the curvature exceeds the Planck scale, the semiclassical approximation (and indeed, spacetime as we know it) will break down. Even in the absence of a final theory of quantum gravity, it is quite remarkable that semiclassical gravity is able to provide such a coherent and elegant picture of black hole evolution in the face of the myriad of instabilities that inner horizons pose.

---